\documentstyle[preprint,eqsecnum,aps,epsf]{revtex}
\begin{document}
\newcommand{\be}{\begin{equation}}
\newcommand{\ee}{\end{equation}}
\newcommand{\zu}{{\bf z}_1}
\newcommand{\zd}{{\bf z}_2}
\newcommand{\ztu}{\dot {\bf z}_{1 {\rm T}}}
\newcommand{\ztd}{\dot {\bf z}_{2 {\rm T}}}

\draft
\preprint{IFUM 499/FT, March 1995}
\title{ Bethe--Salpeter  equation in QCD}

\author{N. Brambilla, E. Montaldi and  G. M. Prosperi}

\address{
Dipartimento di Fisica dell'Universit\`{a} -- Milano\\
INFN, Sezione di Milano -- Via Celoria 16, 20133 Milano
}

\maketitle
\begin{abstract}
We extend to regular QCD the derivation of a confining $ q \bar{q}$
 Bethe--Salpeter equation previously given for the simplest model
 of  scalar QCD in which quarks are treated as spinless particles. We
start  from the same assumptions on the Wilson loop integral already
adopted in the derivation of a semirelativistic heavy quark potential.
 We show that, by standard approximations, an effective meson squared
mass operator can be obtained from our BS kernel and that, from this,
 by ${1\over m^2}$ expansion the corresponding Wilson loop
 potential  can be reobtained, spin--dependent and velocity--dependent
 terms included. We also show that, on the contrary, neglecting
 spin--dependent terms, relativistic flux tube model is reproduced.
\end{abstract}


\section{\bf Introduction}
   In a preceding paper  \cite{scal} we have  derived a confining
Bethe-Salpeter
equation for the simplified model of a scalar
QCD. In that we started from
  the same assumptions
already used in the derivation of the semirelativistic potential for a
heavy quark-antiquark system \cite{BCP}, \cite{BP94}.
The basic object was  the Wilson loop integral
  \be
W = {1\over 3}
 \langle {\rm Tr} {\rm P}_{\Gamma}
 \exp i g \{ \oint_{\Gamma} dx^{\mu} A_{\mu} \}
        \rangle  \ .
\label{eq:loop}
  \ee
where as usual the loop $\Gamma$ is supposed
 made by a quark world line ($ \Gamma _1 $), an
antiquark world line ($ \Gamma _2 $) followed in the
reverse direction and closed by two
straight lines connecting the initial and the final points of the two world
lines ($y_1, y_2$ and $x_1 , x_2$); $A_\mu (x)$  denotes a colour
matrix of the form $A_\mu (x) = {1\over 2} A_\mu^a \lambda^a $;
 ${\rm P}_\Gamma $ prescribes the ordering along the loop
 and ${\rm Tr}$  denotes
the trace of the above
  matrices;
 the expectation value  stands
for the functional  integration
on the gauge field alone.

The basic  assumption was
 \be
i \ln W = i (\ln W)_{\rm pert} + \sigma S_{\rm min}   \,
\label{eq:iniz}
   \ee
$(\ln W)_{\rm pert}$ being the perturbative contribution to $\ln W$ and
$S_{\rm min}$ the minimum area enclosed by $\Gamma$. Furthermore, we used
for $S_{\rm min}$ the straight line approximation, consisting in
replacing
$S_{\rm min}$ with the surface spanned by the straight lines connecting
equal time points on the quark and the antiquark worldlines. In
 practice we
wrote
   \begin{equation}
       S_{\rm min} \cong   \int_{t_{\rm i}}^{t_{\rm f}}
 dt\,  r \int_0^1 ds [1-(s { d {\bf z}_{1 {\rm T}} \over d t}
           + (1-s) {d {\bf z}_{2 {\rm T}} \over d t})^2 ]^{1 \over 2},
\label{eq:minform1}
        \end{equation}
where $t$ stands for the ordinary time in the center of mass frame, ${\bf z}
_1 (t)$ and ${\bf z}_2 (t)$ for the quark and the antiquark positions at the
time $t$, ${dz_{j{\rm T}}^h \over dt}$
denotes the transverse velocity
  $(\delta^{hk}-{r^h r^k \over
r^2}) {dz^k_j \over dt}$ of the particle
 and ${\bf r}(t)$ the relative position ${\bf z}_1(t)-{\bf
z}_2(t)$.\par
 In ref.[1] an essential tool  was the Feynman--Schwinger
path--integral representation for  the spinless
 one--particle propagator
 in an external field.\par
    In this paper we want to extend the
derivation to the case of the  regular QCD with the  lagrangian
\be
L= \sum_{f=1}^{N_f}\bar{\psi}_f (i \gamma^\mu D_\mu -m_f) \psi_f  - {1\over 4 }
 F_{\mu\nu} F^{\mu \nu} + L_{\rm GF}
\label{eq:lagqcd}
\ee
(where $D_{\mu}= \partial_{\mu} -i g A_{\mu}$ and
 $L_{\rm GF}$ is the gauge fixing term),
in which the quarks are fermions
and have spin and $f$ is the flavour index.\par
To do this we find  convenient to work in the second order
formalism.\par
As usual the gauge invariant
 quark--antiquark Green function is given by
\begin{eqnarray}
G^{gi}_4(x_1,x_2,y_1,y_2) &=&
\frac{1}{3}\langle0|{\rm T}\psi_2^c(x_2)U(x_2,x_1)\psi_1(x_1)
\overline{\psi}_1(y_1)U(y_1,y_2)  \overline{\psi}_2^c(y_2)
|0\rangle =
\nonumber\\
&=& \frac{1}{3} {\rm Tr} \langle U(x_2,x_1)
 S_1(x_1,y_1;A) U(y_1,y_2) C^{-1}
S_2(y_2,x_2;A) C \rangle
\label{eq:gauginv}
\end{eqnarray}
where $c$ denotes the charge-conjugate fields, $C$ is the charge-conjugation
matrix, $U$
the path-ordered gauge string
\begin{equation}
U(b,a)= {\rm P}_{ba}  \exp  \left(ig\int_a^b dx^{\mu} \, A_{\mu}(x) \right)
\label{eq:col}
\end{equation}
(the integration path being the straight line joining $a$ to $b$),
 $S_1$ and $S_2$ the quark propagators in the
external gauge field $A^{\mu}$
and obviously
\begin{equation}
\langle f[A] \rangle = \frac{\int {\cal D}[A] M_f(A)
f[A] e^{iS[A]}}
{\int {\cal D}[A] M_f(A) e^{iS[A]}} \> ,
\label{eq:med}
\end{equation}
$S[A]$ being the pure gauge field action and $M_f(A)$
the determinant resulting from the explicit integration on the
fermionic fields (which however in practice we set equal to 1 in
 the adopted approximation). \par
The  propagators  $S_1$ and $S_2$ are supposed to be defined
 by  the equation
\begin{equation}
( i\gamma^\mu D_\mu -m) S(x,y;A) =\delta^4(x-y)
\label{eq:propdir}
\end{equation}
and the appropriate  boundary conditions.\par
Alternatively we can set
\be
S(x,y;A) = (i \gamma^\nu D_\nu + m) \Delta^\sigma(x,y;A)
\ee
and have
\be
(D_\mu D^\mu +m^2 -{1\over 2} g \, \sigma^{\mu \nu} F_{\mu \nu})
\Delta^\sigma (x,y;A) = -\delta^4(x-y)
\label{eq:propk}
\ee
($\sigma^{\mu \nu} = {i\over 2} [\gamma^\mu, \gamma^\nu]$). Then,
 taking into account gauge invariance we can write
\be
G_4^{\rm gi}(x_1,x_2; y_1,y_2) =(i \gamma_1^\mu \partial_{x_1 \mu}
 + m_1) ( i \gamma_2^\nu \partial_{x_2 \nu} +m_2) H_4(x_1,x_2;y_1,y_2)
\label{eq:eqg}
\ee
with
\begin{equation}
H_4(x_1,x_2;y_1,y_2) = -{1\over 3} {\rm Tr}
\langle U(x_2,x_1) \Delta^\sigma_1(x_1,y_1;A)
 U(y_1,y_2) \tilde{\Delta}^\sigma_2(x_2,y_2;-\tilde{A})\rangle
\label{eq:eqh}
\end{equation}
and  the tilde denotes  transposition  on the colour indices alone.
\indent
What   we shall show is
 that the "second order" Green function $H_4(x_1,\,x_2;\,y_1,
\,y_2)$ satisfies a Bethe-Salpeter type nonhomogeneous equation of the form
   \begin{eqnarray}
    H_4(x_1,\,x_2;\,y_1,\,y_2)& \, = &\, H_2(x_1-y_1)\, H_2(x_2-y_2)
   \,- \,i \int d^4\xi_1 d^4\xi_2 d^4\eta_1 d^4\eta_2 \nonumber \\
 &  & H_2(x_1-\xi_1)\, H_2(x_2-\xi_2)\, I_4(\xi_1,\,\xi_2;\,\eta_1,\,\eta_2) \,
    H_4(\eta_1,\,\eta_2;\,y_1,\,y_2) \;
\label{eq:bsh}
\end{eqnarray}
where $H_2$ stands for a kind of colour independent
 one particle dressed propagator
 and $I_4$ denotes the appropriate kernel which is obtained
 as an expansion in the strong coupling constant $\alpha_s =
 {g^2 \over 4 \pi}$ and in the string tension $\sigma$ (better
 in $\sigma a^2$, $a$ being  a characteristic length, typically
 the radius of the particular bound state).\par
At the lowest order in $\alpha_s$ and $\sigma a^2$
  we can write in the momentum space
 \begin{equation}
     \hat{I} (p_1^\prime,\,p_2^\prime;\,p_1,\,p_2)\,=\,\hat I_{\rm pert }
     (p_1^\prime,\,p_2^\prime;\,p_1,\,p_2)\,+\,\hat I_{\rm conf} (p_1^\prime,
     \,p_2^\prime;\,p_1,\,p_2)
\label{eq:bform}
 \end{equation}
$(p_1^\prime +p_2^\prime =p_1 +p_2)$
with
 \begin{eqnarray}
  & & \hat{I}_{\rm pert} =  16 \pi {4 \over 3} \alpha_s
   \{ D_{\rho \sigma}(Q) q_1^\rho
        q_2^\sigma - {i \over 4} \sigma_1^{\mu \nu} (\delta_\mu^\rho Q_\nu-
     \delta_\nu^\rho Q_\mu) q_2^\sigma D_{\rho \sigma } (Q) \nonumber\\
& & + {i \over 4} \sigma_2^{\mu \nu}
        (\delta_\mu^\sigma Q_\nu - \delta_\nu^\sigma Q_\mu) q_1^\rho
  D_{\rho \sigma }(Q)
+{1 \over 16} \sigma_1^{\mu_1 \nu_1} \sigma_2^{\mu_2 \nu_2}
        (\delta_{\mu_1}^\rho Q_{\nu_1} - \delta_{\nu_1}^\rho Q_{\mu_1})
        (\delta_{\mu_2}^\sigma Q_{\nu_2} - \delta_{\nu_2}^\sigma
Q_{\mu_2})
        D_{\rho \sigma}(Q) \}  +\dots\nonumber \\
& &
\label{eq:ipertq}
\end{eqnarray}

\par
   and
  \begin{equation}
     \hat{I}_{\rm conf} = \int d^3 {\bf r} \, e^{i {\bf Q} \cdot {\bf r}}\,
     J({\bf r}, \, q_1, \, q_2)
\label{eq:iconf}
\end{equation}
with
 \begin{eqnarray}
   J({\bf r}, \, q_1, \, q_2) & & = {2 \sigma r \over q_{10} + q_{20} }
    \Big [  q_{20}^2 \sqrt{q_{10}^2  -{\bf q}_{\rm T}^2} +
      q_{10}^2 \sqrt  {q_{20}^2 - {\bf q}_{\rm T}^2} + \nonumber \\
     & &+ {q_{10}^2 q_{20}^2 \over \vert {\bf q}_{\rm T} \vert }
      (\arcsin {\vert {\bf q}_{\rm T}\vert \over  q_{10} } +
      \arcsin {\vert {\bf q}_{\rm T}\vert \over  q_{20}  })
\Big ]
+\nonumber \\
& & +2 \sigma  {\sigma^{k \nu}_1 q_{20} q_{1 \nu} r^k \over r \sqrt{q_{10}^2
+
 {\bf q}_{ {\rm T}}^2}} + 2 \sigma
 { \sigma^{k \nu}_2 q_{10} q_{2 \nu} r^k
\over  r \sqrt{q_{20}^2 - {\bf q}^2_{ {\rm T}}} }
+\dots
\label{eq:iconfj}
\end{eqnarray}
In (1.15)--(1.17) we have set
  \begin{equation}
    q_1={p_1^\prime + p_1 \over 2} \, , \ \  \ \ q_2={p_2^\prime
      + p_2 \over 2} \,  , \
 \  \  \
      Q = p_1^\prime - p_1   = p_2 - p_2^\prime \,  ,
\label{eq:defq}
\end{equation}
 $D_{\rho \sigma}(Q)$ denotes the gluon free
propagator and
the center of mass system is understood
 $ {\bf q}_1 = -{\bf q}_2 = {\bf q} \, , \ q_{\rm T}^h =
(\delta^{hk} - \hat r ^h \hat r ^k) q^k $.\par
   Eqs. (\ref{eq:bsh})-(\ref{eq:iconfj})
 are the basic results of this paper.\par
Notice that
 Eq. (\ref{eq:ipertq}) corresponds to the
usual ladder approximation in this second order  formalism, while
(\ref{eq:iconfj}) reduces to Eq. (1.8) of \cite{scal} when the spin
 dependent terms are neglected.
 Notice also that
instead of (\ref{eq:bsh}) one could have written
the homogenous equation
 \begin{equation}
     \Phi_P (k^\prime) = -i \int {d^4 k \over (2 \pi)^2} \hat H_2(\eta_1 P
+ k^\prime) \hat H_2(\eta_2 P - k^\prime) \hat I(k^\prime , k;P) \Phi_P
(k)
\label{eq:fi}
\end{equation}
which is more appropriate for the bound state problem. In this equation $
\eta_j = {m_j \over m_1 + m_2} $, $ P $ denotes
 the total momentum $p_1 + p_2$,
$ k$  the relative momentum $\eta_2 p_1 - \eta_1 p_2 $
($q_j = \eta_j P + { k+k^\prime \over 2}$ and in the CM frame $ {\bf
q} = { {\bf k}^\prime + {\bf k} \over 2}$ ), $\Phi_P(k) $ is
 the ordinary Bethe--Salpeter wave function in the momentum space.
\par
{}From (\ref{eq:fi}) by replacing $\hat{H}_2 (p)$  with the free propagator
 $ {-i\over p^2 -m^2}  $ and performing an appropriate instantaneous
 approximation on $\hat{I}$ [consisting in setting
$ Q_0=0$, $q_{j0}= { w_j^\prime + w_j\over 2} $ or $
p_{j0}=p_{j0}^\prime ={ w_j^\prime + w_j\over 2} $ or
 $ k_0=k_0^\prime =\eta_2 { w_1^\prime + w_1 \over 2} -\eta_1
{w_2^\prime + w_2\over 2}$ and $ P_0 ={1\over 2} ( w_1^\prime
+ w_1 + w_2^\prime + w_2 )$; with $w_j=\sqrt{m_j^2 + {\bf k}^2 }$,
 $ w_j^\prime=\sqrt{m_j^2 + {\bf k}^{2 \prime}}$ ]
  one can obtain an
effective square mass operator for the mesons (in the CM frame ${\bf
P}=0, P=(m_B,0) $)
  \begin{equation}
    M^2 = M_0^2 + U
\label{eq:quadr}
  \end{equation}
with  $ M_0 = \sqrt{m_1^2 + {\bf k}^2} + \sqrt{m_2^2 + {\bf k}^2} $ and
\begin{equation}
  \langle {\bf k}^\prime \vert U \vert {\bf k} \rangle =
{1\over (2 \pi)^3 }
 \sqrt{ w_1^\prime
  + w_2^\prime \over 2  w_1^\prime  w_2^\prime}\, \hat I_{\rm inst}(
{\bf k}^\prime , {\bf k})  \sqrt{ w_1 + w_2 \over 2  w_1 w_2}.
\label{eq:quadrrel}
\end{equation}
 The quadratic form of Eq.(\ref{eq:quadr}),
 obviously derives from the second order character of the formalism
we have used. It should be mentioned that for light mesons this form
 seems phenomenologically favoured with respect to the linear one.\par
In more usual terms one can also write
\begin{equation}
        M = M_0 + V
\label{eq:lin}
\end{equation}
with
\begin{equation}
\langle {\bf k}^\prime \vert V \vert {\bf k} \rangle =
{1 \over ( 2 \pi)^3 }
 {1 \over 4 \sqrt{ w_1 w_2 w_1^\prime w_2^\prime } } I_{\rm inst}
({\bf k}^\prime, {\bf k})+\dots
\label{eq:linrel}
\end{equation}
where the dots stand for higher order terms in $\alpha _{\rm s}$ and $\sigma
a^2$  and kinematical factors equal to 1 on the energy shell have been
 neglected.
In the limit of small ${{\bf p}^2 \over m^2}$ the potential $V$ as
given by (\ref{eq:quadr})--(\ref{eq:lin})
 reproduces the semirelativistic potential of
 ref. \cite{BCP}, \cite{BP94}.
 Similarly, if we neglect in $V$ the spin dependent terms and the
coulombian one, we
 reobtain the hamiltonian of the relativistic flux tube
model \cite{flux}
 with an appropriate ordering prescription \cite{BP94}. However
we have not yet completely understood the relation
between
the spin
dependent terms we obtain
 and those
appearing in the relativistic flux tube model with ``fermionic ends''
recently proposed \cite{olspin}.\par
     Finally we want to  mention that a result directly
 in hamiltonian form (\ref{eq:lin})
but  strictly
 related to our one has been obtained by
Simonov and collaborators \cite{com,sim}.\par
The paper  is organized in this way.
     In Sect. II we discuss the Feynman--Schwinger
 representation for the one quark
 propagator in an external field,
in Sect. III and IV we study the corresponding representation
 for $H_4$ and
derive the BS equation for such quantity,
 in Sect. V we introduce the
effective mass operator and consider its semirelativistic
limit and its
 relation with the flux tube model.
 Finally in Sect. VI we summarize the results
and make some
additional remarks.

\vskip 1 truecm

\section{  THE FEYNMAN-SCHWINGER REPRESENTATION}

   The solution of Eq.(\ref{eq:propk})
 can be expressed in terms of path integral as
(Feynman-Schwinger representation)
\begin{eqnarray}
& & \Delta^\sigma (x,y;A)  =
- {i\over 2}\int_0^\infty ds \exp {i s\over 2}
 ( - D_\mu D^\mu - m^2 + {1\over 2} g \sigma^{\mu \nu} F_{\mu \nu})
\nonumber \\
& & = -{i\over 2} \int_0^\infty d s
 \int_y^x {\cal D}z\,
  {\rm P}_{xy} {\rm T}_{xy} {\rm exp}\, i \int_0^{s}
 d\tau \{-{1\over 2} (m^2 +\dot{z}^2) + g A_\rho (z) \dot{z}^\rho
 + {g\over 4} \sigma^{\mu \nu} F_{\mu \nu}(z) \}
\label{eq:part}
\end{eqnarray}
where the path integral is understood to be extended over all
 paths
 $z^{\mu}= z^{\mu}(\tau)$ connecting $y$ with $x$ and
expressed in
terms of a   parameter $ \tau$
with $ 0\le \tau
\le s$,
$\dot{z}$  stands
for ${dz(\tau)
\over d \tau}$, the ``functional measure'' is assumed to be
defined as
\be
 {\cal D} z   =  ({1\over 2 \pi i \varepsilon })^{ 2 N}
  d^4 z_1 \dots  d^4 z_{N-1},
\label{eq:defmis}
\ee
   ${ \rm P}_{xy}$ and ${\rm T}_{xy}$  prescribe the ordering
along the path from right to left
respectively of the colour matrices  and of the spin matrices.
\par
On the other side, it is well known
that, as a consequence of
 a variation in the path $z^\mu(\tau)
\to z^\mu(\tau)+ \delta z^\mu(\tau)$
respecting the extreme points, one finds
\begin{eqnarray}
\delta & &\big \{ {\rm P}_{xy} \exp ig \int_0^s d\tau \dot{z}^\mu(\tau) A_{\mu}
(z)\big \}=\nonumber \\
& & = ig \int_0^s \delta S^{\mu\nu}(z(\tau)) {\rm P}_{xy}
\big \{- F_{\mu \nu}(z(\tau)) \exp ig \int_0^s d\tau^\prime
\dot{z}^\mu(\tau^\prime) A_{\mu}(z(\tau^\prime)) \big \}
\label{eq:varwils}
\end{eqnarray}
with $\delta S^{\mu \nu}(z)= {1\over 2} (d z^\mu \delta z^\nu- dz^\nu
\delta z^\mu)$.
 So one  can  naturally  write
\begin{eqnarray}
{\rm T}_{xy}& &  \exp(-{1\over 4} \int_0^s d\tau \sigma^{\mu\nu} {\delta
\over \delta S^{\mu \nu}(z) } )
\Big ( {\rm P}_{xy} \exp ig \int_0^s d \tau^\prime \dot{z}^\mu(\tau^\prime)
A_{\mu}(z(\tau^\prime)) \Big )\nonumber \\
& & = {\rm T}_{xy} {\rm P}_{xy} \exp ig \int_0^s d\tau [
\dot{z}^\mu(\tau)A_{\mu}(z(\tau)) + {1\over 4} \sigma^{\mu \nu}
F_{\mu \nu}(z(\tau )) ]
\label{eq:deffmu}
\end{eqnarray}
 and
Eq.(\ref{eq:part}) can  be rewritten as
\be
\Delta^\sigma(x,y; A)= -{i\over 2} \int_0^s d\tau \int_y^x {\cal
D}z  {\rm P}_{x y}
{\rm T}_{xy} {\cal S}_0^s \exp \, i \int_0^s d\tau [-{1\over 2} (m^2
+\dot{z}^2)
+ ig\dot{\bar{z}}^\mu A_{\mu}(\bar{z})]
\label{eq:partbis}
\ee
with
\be
{\cal S}_0^s =  \exp \Big [ -{1\over 4} \int_0^s d \tau
 \sigma^{\mu \nu} {\delta \over \delta S^{\mu \nu}(\bar{z})}
\Big ]
\label{eq:defop}
\ee
In (2.6)
 it is understood that $ \bar{z}^\mu(\tau)$ has to be put
equal to  $z^\mu(\tau)$ after the action of ${\cal S}_0^{s_1} $.
 Alternatively, it is convenient
to write $\bar{z}= z+\zeta$, assume that
${\cal S}_0^s$ acts on $\zeta (\tau)$ with
$ \delta S^{\mu \nu} (z) = {1\over 2} ( d z^\mu \delta \zeta^\nu
 - d z^\nu \delta \zeta^\mu )$ and set eventually $\zeta =0$.

\section{  THE FEYNMAN--SCHWINGER REPRESENTATION}

 Replacing (\ref{eq:partbis}) in (\ref{eq:eqh}) we obtain
\begin{eqnarray}
H_4(x_1,x_2;y_1,y_2) & & = ({1 \over 2})^2 \int_0^{\infty} d s_1
\int_0^{\infty} d s_2
 \int_{y_1}^{x_1} {\cal D} z_1\int_{y_2}^{x_2} {\cal D} z_2
 {\rm T}_{x_1 y_1} {\rm T}_{x_2 y_2}
 {\cal S}_0^{s_1} {\cal S}_0^{s_2} \nonumber \\
& & \exp{ ({-i \over 2}) \big \{
 \int_0^{s_1} d\tau_1 (m_1^2 +\dot{z}_1^2) + \int_0^{s_2}
d\tau_2 (m_2^2 +\dot{z}_2^2)\big \} }\nonumber \\
& & {1\over 3} \langle {\rm Tr} {\rm P}_\Gamma
 \exp (ig) \big \{ \oint_{\Gamma} d\bar{z}^\mu A_{\mu}
(\bar{z})   \}
\rangle
\label{eq:hquatr}
\end{eqnarray}
where now $ \bar{z}= \bar{z}_j= z_j +\zeta_j $ on $\Gamma_1 $ and
$\Gamma_2$, $ \bar{z}=z $ on the end lines $x_1 x_2$ and $y_2 y_1$
 and the final limit $\zeta_j \to 0$ is again  understood.\par
Then, let us try to be more explicit concerning Eq. (1.2) and (1.3).
For the first term in (1.2) we have,
 at the lowest order of perturbation theory,
\begin{eqnarray}
& &i (\ln W)_{\rm pert} = i
 \ln \langle {1\over 3} {\rm Tr} {\rm P} \exp ig \oint_{\Gamma} dz^{\mu}
 A_{\mu}(z)\rangle_{\rm pert} ={4\over 3} g^2 \int_0^{s_1} d \tau_1
\int_0^{s_2} d\tau_2 D_{\mu \nu}(z_1-z_2)
 \dot{z}_1^{\mu}
\dot{z}_2^{\nu} \nonumber \\
& & -{2\over 3} g^2 \int_0^{s_1} d \tau_1
\int_0^{s_1} d\tau_1^{\prime}
 D_{\mu \nu}(z_1-z_1^{\prime}) \dot{z}_1^{\mu}
\dot{z}_1^{\prime\nu}-
{2\over 3} g^2 \int_0^{s_2} d \tau_2
\int_0^{s_2} d\tau_2^{\prime} D_{\mu
\nu}(z_2-z_2^{\prime}) \dot{z}_2^{\mu}
\dot{z}_2^{\prime \nu}+ \dots
\label{eq:wilpert}
\end{eqnarray}
On the other side, for the second one in general we have to write
\begin{equation}
S_{\rm min} = \int_{t_i}^{t_f} dt
 \int_0^1 ds \big [ -({\partial u^\mu \over \partial t} {\partial u_\mu
\over \partial t })({\partial u^\mu \over \partial s} {\partial u_\mu
\over \partial s })+
 ({\partial u^\mu \over \partial t} {\partial u_\mu
\over \partial s})^2 \big ]^{1\over 2}
\label{eq:smin}
\end{equation}
$x^\mu = u^\mu (t,s) $ being  the equation of the minimal surface
with  contour $\Gamma$. Let us  assume  that for fixed $t$ and for $s$
varying from 0 to 1, $u^\mu (s,t)$ describes a line connecting a point
on the quark world line $\Gamma_1$  with one on the antiquark
 world line $\Gamma_2$,
\begin{equation}
u^\mu(1,t)= z_1^\mu(\tau_1(t)),\quad \quad \quad
u^\mu(0,t)= z_2^\mu(\tau_2(t))
\label{eq:udef}
\end{equation}
Obviously (\ref{eq:smin}) is invariant under reparametrization, so
 a priori $t$ and $s$ could be everything.
In particular if  $\Gamma_1$ and $\Gamma_2$
never go backwards in time,
  $t$  can be choosen
as the ordinary time $u^0(s,t)\equiv t$.
For the moment let us assume  that this is the case.
Then
$\tau_1 (t)$ and $\tau_2(t)$ are specified by the equation
\begin{equation}
z_1^0(\tau_1) =z_2^0 (\tau_2)
\label{eq:zzero}
\end{equation}
and we can  set
\begin{equation}
L=\int_0^1 ds
\big [ -({\partial u^\mu \over \partial t} {\partial u_\mu
\over \partial t })({\partial u^\mu \over \partial s} {\partial u_\mu
\over \partial s })+
 ({\partial u^\mu \over \partial t} {\partial u_\mu
\over \partial s})^2 \big ]^{1\over 2}.
\label{eq:defl}
\end{equation}
Obviously  $L$  cannot depend only on
 on the extremal points
 $z_1(\tau_1)$ and $z_2(\tau_2)$ but has to depend
 even on the shape of the
 world lines at least in a neighbourhood of such points. So,
we can think of it as  a function of all derivatives in $\tau_1$
 and $\tau_2$ and write
$L=L(z_1,z_2, \dot{z}_1,
\dot{z}_2, \dots )$.
Finally (3.3) can be rewritten as
\begin{equation}
S_{\rm min}= \int dz_1^0 \int dz_2^0 \delta (z_1^0-z_2^0)
L(z_1,z_2,\dot{z}_1,\dot{z}_2, \dots)=
\int d\tau_1 \int d\tau_2 \delta(z_1^0-z_2^0) \dot{z}_1^0 \dot{z}_2^0
L(z_1,z_2,\dot{z}_1,\dot{z}_2, \dots ).
\label{eq:deflmin}
\end{equation}
 and
 in principle this expression can be
considered a good approximation
  even if
 the world lines contain pieces going backwards in time. In fact,
in such a case
  if we fix e.g. $\tau_1$,
 (\ref{eq:zzero})  has more than one solution in
$\tau_2$ and if $\Gamma_1$ and $\Gamma_2$ are not too much irregular
in space ( otherwise $S_{\rm min}$ is large and the weight of the loop
 is small)
the minimal surface can be reconstructed as the algebraic
 sum of various pieces of surface.\par
In  the straight line approximation we must choose
\begin{eqnarray}
u^0(s,t) & = & t \nonumber \\
u^k(s,t) & = & s z_1^k (\tau_1(t)) + (1-s) z_1^k(\tau_2(t))
\label{eq:strai}
 \end{eqnarray}
and
 we have
\be
 \dot{z}_1^0 \dot{z}_2^0 L=  \sigma
  \vert {\bf z}_1^{\prime} -{\bf z}_2^{\prime}\vert
\int_0^1 d s \big \{ {\dot{z}_{10}^{\prime 2}} {\dot{z}_{20}^{\prime 2}} -
 (s \dot{\bf z}_{1{\rm T}}^{\prime} \dot{z}_{20}^{\prime }
 + (1-s) \dot{\bf z}_{2 {\rm T}}^{\prime}
 \dot{z}_{10}^{\prime } )^2 \big \}^{1\over 2}
\label{eq:lstrai}
\ee
which introduced in (3.7) becomes  equivalent to Eq.(1.3).
The important point concerning  (3.7) with (3.9) is that it has the
same general  form as (3.2).
However we stress that the  approximation (3.8) must
  be  performed only {\it
after} that the application of the operators ${\cal S}_0^{s_1}$
and ${\cal S}_0^{s_2}$ has been performed.\par
Substituting (\ref{eq:wilpert}) and (3.7) in
(\ref{eq:hquatr})
 we obtain
\begin{eqnarray}
& & H_4(x_1,x_2;y_1,y_2)  =  ({1 \over 2})^2 \int_0^{\infty} ds_1
\int_0^{\infty} ds_2
 \int_{y_1}^{x_1} {\cal D} z_1\int_{y_2}^{x_2} {\cal D} z_2
{\rm T}_{x_1 y_1} {\rm T}_{x_2 y_2}
 {\cal S}_0^{s_1} {\cal S}_0^{s_2} \nonumber \\
& & \exp i  \big \{ -{1\over 2}
 \int_0^{s_1} d\tau_1 (m_1^2 +\dot{z}_1^2) -{1\over 2} \int_0^{s_2}
d\tau_2 (m_2^2 +\dot{z}_2^2)+ \nonumber \\
& &  +{2\over 3} g^2 \int_0^{s_1} d\tau_1
\int_0^{s_2} d\tau_1^\prime D_{\mu \nu}(\bar{z}_1-\bar{z}_1^\prime)
\dot{\bar{z}}_1^\mu \dot{\bar{z}}_1^{\nu\prime}+
{2\over 3} g^2
\int_0^{s_2} d\tau_2 \int_0^{s_2} d\tau_2^\prime D_{\mu \nu}(\bar{z}_2
-\bar{z}_2^\prime) \dot{\bar{z}}_2^\mu \dot{\bar{z}}_2^{\nu\prime}
\nonumber \\
& &
- \int_0^{s_1} d\tau_1 \int_0^{s_2} d\tau_2
E(\bar{z}_1, \bar{z}_2, \dot{\bar{z}}_1, \dot{\bar{z}}_2,\dots)\big
\},
\label{eq:hpath}
\end{eqnarray}
where we have set
\begin{eqnarray}
E(z_1, z_2, \dot{z}_1, \dot{z}_2 \dots ) = & & {4\over 3} g^2
 D_{\mu \nu} (z_1-z_2) \dot{z}_1^\mu \dot{z}_2^\nu + \nonumber \\
 & & + \sigma \delta(z_{10} - z_{20}) \dot{z}_{10} \dot{z}_{20}
 L(z_1, z_2, \dot{z}_1, \dot{z}_2 \dots ).
\label{eq:defe}
\end{eqnarray}
Now let
 us  denote   the quantity in
 curly bracket in (3.10)  by $S_4$
 and
  perform a Legendre transformation by introducing the momenta
  $ p_{j \mu}=-
{\delta S_4^{q \bar{q}}\over \delta \dot{z}_{j}^{\mu}}$
( in this the various quantities $z_j$, $\dot{z}_j$,
 $\ddot{z}_j, \dots $ are assumed to be treated as
 independent)
\begin{eqnarray}
 { p}_{\mu 1}
 & =& \dot{{ z}}_{\mu 1} + {4\over 3} g^2\int_0^{s_1}
 d\tau_1^\prime D_{\mu \nu}(\bar{z}_1-\bar{z}_1^{\prime})
\dot{\bar{z}}_1^{\prime\nu} + \int_0^{s_2} d\tau_2^\prime {
\partial E( \bar{z}_1, \bar{z}_2^\prime. \dot{\bar{z}}_1 ,
\dot{\bar{z}}_2 \dots ) \over \partial \dot{z}_1^\mu}
\nonumber\\
 {p}_{\mu 2} & =& \dot{z}_{\mu 2} +{4\over 3} g^2 \int_0^{s_2}
 d\tau_2^\prime D_{\mu \nu}(z_2-z_2^{\prime})
\dot{\bar{z}}_2^{\prime\nu}+
 \int_0^{s_1} d\tau_1^\prime {
\partial E( \bar{z}_1, \bar{z}_2^\prime. \dot{\bar{z}}_1 ,
\dot{\bar{z}}_2 \dots ) \over \partial \dot{z}_2^\mu}.
\label{eq:momqq}
\end{eqnarray}
Eq.(\ref{eq:momqq}) cannot be inverted in a closed form
 with respect to $\dot{z}_1$ and $\dot{z}_2$.
 However, we can
do this by an expansion in $\alpha_s= {g^2 \over 4 \pi }$ and
${\sigma a^2}$ and at the lowest order we have
\begin{eqnarray}
\dot{z}_1^{\mu}& =&  p_1^{ \mu} -
{4\over 3} g^2 \int_0^{s_1} d \tau_1^\prime  D_{\mu
\nu}(\bar{z}_1 -z_1^{\prime}) \bar{p}_1^{\prime \nu}
- \int_0^{s_2} d\tau_2^\prime {
\partial E( \bar{z}_1, \bar{z}_2^\prime, \bar{p}_1 ,
\bar{p}_2^\prime  \dots ) \over \partial p_1^\mu}
+\dots
\nonumber \\
\dot{z}_2^{\mu}& =&  p_2^{\mu}
-{4\over 3} g^2 \int_0^{s_2} d \tau_2^\prime D_{\mu
\nu}(\bar{z}_2- \bar{z}_2^{\prime}) p_2^{\prime \nu}-
 \int_0^{s_1} d\tau_1^\prime {
\partial E( \bar{z}_1, \bar{z}_2^\prime, \bar{p}_1^\prime ,
\bar{p}_2  \dots ) \over \partial p_2^\mu}
+\dots
\label{eq:defv}
\end{eqnarray}
with
\begin{equation}
\bar{p}_j^\mu = p_j^\mu + \dot{\zeta}_j^\mu .
\label{eq:defpvar}
\end{equation}
In conclusion we find
\begin{eqnarray}
 & & \quad \quad H_4(x_1,x_2,y_1,y_2) = ({1 \over 2})^2 \int_0^\infty ds_1
\int_0^\infty ds_2 \int_{y_1}^{x_1} {\cal D}z_1 {\cal D}p_1
 \int_{y_2}^{x_2} {\cal D}z_2 {\cal D}p_2 T_{x_1 y_1} T_{x_2 y_2}
\nonumber \\
& &  {\cal S}_0^{s_1} {\cal S}_0^{s_2}
  \exp i \Big \{ \int_0^{s_1} d\tau_1 K_1 + \int_0^{s_2} d\tau_2
 K_2   -  \int_0^{s_1} d\tau_1 \int_0^{s_2} d\tau_2
E(\bar{z}_1, \bar{z}_2, \bar{p}_1, \bar{p}_2, \dots )+\dots \Big \},
\label{eq:defhe}
\end{eqnarray}
where
\begin{equation}
K_j= -p_j\cdot \dot{z}_j +{1\over 2} (p_j^2 -m_j^2)
+{2\over 3} g^2 \int_0^{s_j} d\tau_j^\prime D_{\mu \nu} (\bar{z}_j
-\bar{z}_j^\prime) \bar{p}_j^\mu \bar{p}_j^\nu
\label{eq:defk}
\end{equation}
includes the self--interacting term.  Notice that now in ${\cal
S}_0^{s_j}$ it  must be understood  $ \delta S^{\mu \nu}
(\bar{z}_j) ={1\over 2} d \tau_j (p_j^\mu \delta \zeta_j^\nu -p_j^\nu
\delta \zeta_j^\mu ) +\dots$. \par
Eq. (\ref{eq:defhe})
 is the starting point for the derivation
 of our Bethe--Salpeter equation.

\section{THE HOMOGENEOUS BETHE--SALPETER EQUATION}
In  Eq. (\ref{eq:defhe}) we proceed along the same line followed in Ref.
\cite{scal}. \par
Applying to the interaction term $E$  the identity
\begin{equation}
\exp \int_0^s d \tau A(\tau) = 1 + \int_0^s d \tau A(\tau) \exp
 \big ( \int_0^\tau d \tau^\prime A(\tau^\prime) \big )
\label{eq:id}
\end{equation}
 we have
\begin{eqnarray}
& & H_4(x_1,x_2;y_1,y_2) = ({1 \over 2})^2
\int_0^\infty ds_1
\int_0^\infty ds_2 \int_{y_1}^{x_1} {\cal D}z_1 {\cal D}p_1
 \int_{y_2}^{x_2} {\cal D}z_2 {\cal D}p_2
\nonumber \\
& & T_{x_1 y_1} T_{x_2 y_2}  {\cal S}_0^{s_1} {\cal S}_0^{s_2}
 \bigg \{ \exp i [ \int_0^{s_1} d\tau_1 K_1 + \int_0^{s_2} d\tau_2
 K_2 ]- i \int_0^{s_1} d \tau_1 \int_0^{s_2} d\tau_2
 E(\bar{z}_1, \bar{z}_2, \bar{p}_1, \bar{p}_2\dots )
\nonumber \\
& &\times  \exp i \Big \{  \int_0^{s_1} d\tau_1 K_1 + \int_0^{s_2} d\tau_2
 K_2
 - \int_0^{\tau_1} d\tau_1^\prime
\int_0^{s_2} d\tau_2^\prime
 E(\bar{z}_1^\prime, \bar{z}_2^\prime,
\bar{p}_1^\prime, \bar{p}_2^\prime, \dots)  \bigg \}
\label{eq:hlav}
\end{eqnarray}
\indent Now, using the method of Ref. \cite{scal} and having in mind
 (\ref{eq:defe}), (\ref{eq:defl}) and (\ref{eq:hpath})
  one finds (see Appendix for details)
\begin{eqnarray}
 {\delta \over \delta S^{\mu \nu}(z_1^\prime)} & &  \int_0^{s_1} d
\tau_1^\prime \int_0^{s_2} d \tau_2^\prime E(z_1^\prime, z_2^\prime,
p_1^\prime, p_2^\prime, \dots) =\nonumber \\
& & = \int_0^{s_2} d\tau_2^\prime \Big [ {4\over 3} g^2 \big (
\partial_\nu D_{\mu \sigma} (z_1-z_2^\prime) -\partial_{\mu} D_{\nu
\sigma}(z_1-z_2^\prime) \big )p_2^\sigma +\nonumber \\
& & + \sigma \delta(z_{10}-z_{20}) { p_{1\nu} (z_{1\mu}
-z_{2\mu}^\prime ) - p_{1\mu} (z_{1\nu} -z_{2\nu}^\prime )
\over \sqrt{ ({p}_{10}^2 -\dot{\bf p}_1^2 ) ( {\bf z}_1- {\bf z}_2 )^2
+ ( {\bf p}_1 \cdot ({\bf z}_1- {\bf z}_2^\prime))^2 }}
 +\dots  \Big ]
\label{eq:varnonpert}
\end{eqnarray}
and a similar result, with a minus sign of difference in front, for the
derivative in $ {\delta \over \delta S^{\mu \nu} (z_2^\prime)}$.\par
Furthermore
\begin{equation}
{\delta^2 \over   \delta S^{\mu \nu} (z_1)  \delta S^{\rho \sigma}
(z_1^\prime) } \int_0^{s_1} d\tau_1^{\prime \prime} \int_0^{s_2}
 d\tau_2^{\prime \prime} E = {\delta^2 \over \delta S^{\mu \nu}(z_2)
\delta S^{\rho \sigma}(z_2^\prime ) }
\int_0^{s_1} d\tau_1^{\prime \prime} \int_0^{s_2} d\tau_2^{\prime
\prime} E= 0
\label{eq:varzero}
 \end{equation}
but
\begin{equation}
{\delta^2 \over \delta S^{\mu_1 \nu_1} (z_1) \delta S^{\mu_2\nu_2}
(z_2) } \int_0^{s_1} d\tau_1^{\prime \prime} \int_0^{s_2}
d\tau_2^{\prime \prime} E = {4\over 3} g^2
(\delta_{\mu_1}^\rho \partial_{\nu_1} -\delta_{\nu_1}^\rho
\partial_{\mu_1} ) ( \delta_{\mu_2}^\sigma \partial_{\nu_2}
-\delta_{\nu_2}^\sigma \partial_{\mu_2} ) D_{\rho \sigma} (z_1-z_2)
\label{eq:varpert}
\end{equation}
\indent Then,
taking into account the relation
\begin{equation}
e^A B e^{-A} =\sum_{n=0}^\infty {1\over n !} [A,[A, \dots [A,B] \dots
]],
\label{eq:idcom}
\end{equation}
and specifically  (\ref{eq:varzero}), we have
\begin{eqnarray}
 & &
 {\cal S}_0^{s_1}
 {\cal S}_0^{s_2} \int_0^{s_1} d\tau_1 \int_0^{s_2} d\tau_2
    E(\bar{z}_1, \bar{z}_2, \bar{p}_1,
\bar{p}_2, \dots )  ({\cal S}_0^{s_1}
 {\cal S}_0^{s_2} )^{-1}= \nonumber \\
& &= (1-{1\over 4} \int_0^{s_1} d s_1^\prime \sigma_1^{\mu \nu}
{\delta \over \delta S^{\mu_1 \nu_1}(\bar{z}_1^\prime )})
(1-{1\over 4} \int_0^{s_2} d s_2^\prime \sigma_2^{\mu \nu}
{\delta \over \delta S^{\mu_2 \nu_2}(\bar{z}_2^\prime )})
\nonumber \\
& & \int d\tau_1 \int d\tau_2
  E(\bar{z}_1, \bar{z}_2, \bar{p}_1,
\bar{p}_2, \dots )= R(z_1, z_2, p_1,p_2)
\label{eq:defr}
\end{eqnarray}
with
\begin{equation}
R  = R_{\rm pert} + R_{\rm conf}
\end{equation}
\begin{eqnarray}
& & \quad  R_{\rm pert}=   -{4\over 3} g^2  \Big \{
 D_{\rho \sigma} (z_1-z_2) p_1^\rho p_2^\sigma \nonumber \\
& & -{1\over 4} \sigma_1^{\mu \nu} (\delta_{\mu}^{\rho} \partial_{1\nu}
-\delta_\nu^\rho \partial_{1\mu} ) D_{\rho \sigma}(z_1-z_2)p_2^\sigma
- {1\over 4}\sigma_2^{\mu \nu} (\delta_{\mu}^{\sigma} \partial_{2\nu}
-\delta_\nu^\sigma \partial_{2\mu} ) D_{\rho \sigma}(z_1-z_2)p_1^\rho
\nonumber \\
& & +{1\over 16} \sigma_1^{\mu_1 \nu_1} \sigma_2^{\mu_2 \nu_2}
 (\delta_{\mu_1}^\rho \partial_{1\nu_1} -\partial_{\nu_1}^\rho
\partial_{1\mu_1} )
(\delta_{\mu_2^\sigma}\partial_{2\nu_2} -\partial_{\nu_2}^\sigma
\partial_{2\mu_2} )  D_{\rho \sigma}(z_1- z_2)\Big \}
\label{eq:rpert}
\end{eqnarray}
and
\begin{eqnarray}
R_{\rm conf} & & =  \sigma \delta(z_{10}-z_{20}) \bigg \{
\vert {\bf z}_1- {\bf z}_2 \vert \int_0^1  ds \sqrt{p_{10}^2 p_{20}^2
-[s p_{1{\rm T}} p_{20} + (1-s) p_{2 {\rm T}} p_{10} ]^2 }
\nonumber \\
&& -{1\over 4} p_{20} \sigma_1^{\mu\nu} { p_{1\nu} (z_{1\mu}
-z_{2\mu})  -p_{1\mu} ( z_{1\nu} -z_{2\nu}) \over
 \vert {\bf z}_1 -{\bf z}_2 \vert \sqrt{p_{10}^2 -{\bf p}_{1 {\rm
T}}^2 }}
 +{1\over 4} p_{10} \sigma_2^{\mu\nu} { p_{2\nu} (z_{1\mu}
-z_{2\mu}) -p_{2\mu} ( z_{1\nu} -z_{2\nu}) \over
 \vert {\bf z}_1 -{\bf z}_2 \vert \sqrt{p_{20}^2 -{\bf p}_{2 {\rm
T}}^2 }}\bigg \}.
\label{eq:rspez}
\end{eqnarray}
Finally setting
\begin{equation}
H_2(x-y) ={-i\over 2}\int_0^\infty ds
 \int_y^x {\cal D}z {\cal D}p\, {\cal S}_0^s \exp i
\int_0^s d\tau K
\label{eq:hdue}
\end{equation}
Eq.(1.18) can be written
\begin{eqnarray}
& & \quad \quad H_4(x_1,x_2; y_1,y_2)  = H_2 (x_1-y_1) H_2(x_2-y_2)+
\nonumber\\
& & - {i\over 4} \int_0^{\infty} d s_1 \int_0^\infty d s_2
 \int_{y_1}^{x_1} {\cal D}z_1 {\cal D}p_1 \int_{y_2}^{x_2}
 {\cal D} z_2 {\cal D} p_2  {\rm T}_{x_1 y_1} {\rm T}_{x_2 y_2}
 \int_0^{s_1}d\tau_1
 \int_0^{s_2} d\tau_2  R(z_1, z_2, p_1,p_2) \nonumber \\
& & {\cal S}_0^{s_1} {\cal S}_0^{s_2}  \exp i \Big \{ \int_0^{s_1}
 d\tau_1^\prime  K_1^\prime + \int_0^{s_2} d\tau_2^\prime
 K_2^\prime -i \int_0^{\tau_1} d\tau_1^\prime \int_0^{s_2}
 d\tau_2^\prime  E(z_1^\prime, z_2^\prime, p_1, p_2 \dots )\Big \}.
\label{eq:hquatrr}
\end{eqnarray}
\par At this point it is necessary to take explicitely into account the
discrete  form  of (\ref{eq:hquatrr}). If we take
\begin{equation}
{\rm P} \exp  \big [i g \oint_{\Gamma} dz^\mu A_{\mu}(z) \big ]
 = {\rm P}
\prod_{\Gamma}  U(z_n, z_{n-1}) = {\rm P} \exp i g
 \sum_{\Gamma} (z_n^\mu - z_{n-1}^\mu ) A_\mu ( {z_n + z_{n-1}\over 2}
)
\label{eq:wilsdiscr}
\end{equation}
(as required by a gauge invariant  definition of the integral on the
gluon field) we have
\begin{eqnarray}
& & H_4(x_1,x_2; y_1,y_2)  = H_2(x_1-y_1) H_2(x_2-y_2) - {i\over 4}
 \varepsilon^2 \sum_{N_1=0}^\infty \sum_{N_2=0}^{\infty}\nonumber \\
& & {1\over (2\pi )^{N_1+N_2} } \int d^4p_{11} d^4 z_{11} \dots d^4 z_{1
N-1} d^4 p_{1 N_1} \int d^4 p_{21} d^4 z_{21} \dots d^4 z_{2 N_2-1}
 d^4 p_{2 N_2-1} {\rm T}_{x_1 y_1} {\rm T}_{x_2 y_2}\nonumber \\
& & \sum_{R_1 =1}^{N_1-1} \sum_{R_2=1}^{N_2-1}  R({z_{1 R_1} + z_{1
R_1-1}\over
2} , { z_{2 R_2} + z_{ 2 R_2-1} \over 2}, p_{1R}, p_{2R} )
\nonumber \\
& &  {\cal
S}_0^{s_1} {\cal S}_0^{s_2} \exp i \Big \{ \sum_{j=1}^2
\sum_{n=1}^{N_j} \big [
- p_{jn } (z_{jn}- z_{j n-1} ) + \varepsilon (p_{jn}^2 - m_j^2 ) -
\nonumber \\
& & - {4\over 3} g^2 \varepsilon^2 \sum_{n^\prime =1}^{n-1}
D_{\mu \nu} ( { z_{jn^\prime} + z_{j n^\prime -1} \over 2} -
 { z_{j n^\prime } +z_{j n^\prime -1} \over 2} ) p_{j n^\prime}^\mu
p_{j n^\prime}^\nu \big ]\nonumber \\
& & - \varepsilon^2 \sum_{n_1=1}^{R_1 -1} \sum_{n_2=1}^{N_2}
 E( { z_{1n} + z_{1 n-1} \over 2}, {  z_{2n} + z_{2 n-1}\over 2},
p_{1n}, p_{2n}, \dots ) \Big \}
\label{eq:discr1}
\end{eqnarray}
If we neglect in the exponent the ``non planar'' terms
$\sum_{n=R_j+1}^{N_j} \sum_{n^\prime =1}^{R_j}
 D_{\mu \nu} p_{j n_j}^\mu p_{j n_j^\prime}^\nu $
 and $\sum_{n_1=1}^{R_1} \sum_{n_2= R_2+1}^{N_2} D_{\mu \nu} p_{1
n}^\mu
p_{2 n_2}^\nu $, corresponding in the continuous to the quantity
\begin{equation}
\int_{\tau_j}^{s_j} d \tau_j^\prime \int_0^{\tau_j} d\tau_j^{\prime
\prime} D_{\mu \nu} (z_j^\prime -z_j^{\prime \prime} ) p_j^{\mu
\prime} p_j^{\nu \prime \prime}
\label{eq:nonplancont}
\end{equation}
and
\begin{equation}
\int_0^{\tau_1} d\tau_1^\prime \int_{\tau_2}^{s_2} d\tau_2^\prime
D_{\mu \nu} (z_1^\prime - z_2^\prime ) p_1^{\mu \prime} p_2^{\prime
\nu},
\label{eq:nonplancont2}
\end{equation}
Eq.(\ref{eq:discr1}) can be written
\begin{eqnarray}
& & \quad  \quad
H_4(x_1,x_2,y_1,y_2)  = H_2(x_1-y_1) H_2(x_2-y_2) - {i\over 4}
 \varepsilon^4  \sum_{R_1=1}^\infty \sum_{R_2=1}^\infty
 \sum_{N_1=R_1+1}^\infty  \sum_{N_2 = R_2 +1}^\infty \nonumber \\
& & {1\over (2 \pi )^4} \int d^4 z_{1 R_1}  d^4 p_{1 R_1} d^4 z_{1
R_1-1} {1\over (2 \pi )^4 } \int d^4 z_{2 R_2} d^4 p_{2 R_2}
 d^4  z_{R_2 -1} \int_{z_{1 R_1+1 }}^{x_1} {\cal D}z_1 {\cal D} p_1
 \int_{z_{2 R_2 +1}}^{x_2} {\cal D} z_2 {\cal D}p_2 \nonumber \\
 & & \int_{y_1}^{z_{1 R_1-1}} {\cal D} z_1 {\cal D} p_1
\int_{y_2}^{z_{2 R_2-1}} {\cal D }z_2 {\cal D}p_2 \nonumber \\
& &  {\rm T}_{x_1 z_{1R}}
 {\rm T}_{x_2 z_{2 R}} {\cal S}_{\tau_1}^{s_1} {\cal S}_{\tau_2}^{s_2}
\exp i \Big \{ \sum_{j=1}^2 \sum_{n=R_j +1 }^{N_1}
 [- p_{j n} (z_{j n} - z_{j n-1} ) +\varepsilon ( p_{j n}^2 - m_j^2 )
-{4\over 3} g^2 \varepsilon^2 \sum_{n^\prime=R_j+1 }^{N_1}  D_{\mu \nu} p_{j
n}^\mu  p_{j n^\prime}^\nu  ] \Big \}\cdot \nonumber \\
& & \exp \big [ -i \sum_{j} p_{j R} (z_{j R_j} - z_{j R_{j-1}}) \big ]
 R({ z_{1 R_1} + z_{1 R_1-1} \over 2}, {z_{2 R_2} + z_{2 R_2-1}
\over 2}, p_{1R}, p_{2 R} )\cdot \nonumber \\
 & & {\rm T}_{z_{1 R}y_1} {\rm T}_{ z_{2 R} y_2}
 {\cal S}_0^{\tau_1} {\cal S}_0^{\tau_2}
 \exp  i \Big \{ \sum_{j=1}^2 \sum_{n=1}^{R_j-1} \big [- p_{j n}
 ( z_{j n} - z_{ j n-1}  ) + \varepsilon (p_{jn}^2 - m_j^2 )-
\sum_{n^\prime =1 }^{R_j -1} D_{\mu \nu} p_{j n} p_{j n^\prime} \big ]
+\nonumber \\
& & + \sum_{n_1 =1}^{R_1 -1} \sum_{n_2 =1}^{R_2-1} E( { z_{1 n_1} + z_{1
n_1-1} \over 2 } , { z_{2 n_2} + z_{2 n_2-1} \over 2} , p_{1 n_1},
 p_{2 n_2} ) \Big \}
\label{eq:discrdue}
\end{eqnarray}
which, going back to the continuous corresponds to Eq.(\ref{eq:bsh})
with
\begin{equation}
I(\xi_1, \xi_2 , \eta_1, \eta_2 ) = -4 i \int { d^4 k_1 d^4 k_2 \over (2
\pi )^8}  R( {\xi_1 + \eta_1 \over 2}, { \xi_2 +\eta_2 \over 2},
k_1 , k_2 ) \exp \big  \{(-i) [ (\xi_1 -\eta_1 )
 k_1 + (\xi_2 -\eta_2 ) k_2 ]\big \}
\label{eq:kernr}
\end{equation}
In conclusion,
taking the Fourier transform
\begin{eqnarray}
& & \quad \quad \quad (2 \pi )^4 \delta (p_1^\prime +p_2^\prime - p_1 p_2 )
\hat{I} ( p_1^\prime, p_2^\prime ; p_1, p_2) =
  - 4 i\int d^4 \xi_1  d^4 \xi_2 \int d^4 \eta_1 d^4 \eta_2
\nonumber \\
& & \int { d^4 k_1 \over ( 2 \pi )^4 } {d^4 k_2 \over ( 2 \pi )^4 }
e^{ i (p_1^\prime - k_1 ) \xi_1 + i ( p_2^\prime - k_2 ) \xi_2 }
  R( { \xi_1 +\eta_1 \over 2}, { \xi_2 +\eta_2 \over 2}, k_1, k_2 )
 e^{ - i (p_1 -k_1 )\eta_1 -i (p_2-k_2) \eta_2 }
\label{eq:kernp}
\end{eqnarray}
we obtain (1.14)--(\ref{eq:iconfj}).

\section{EFFECTIVE MASS OPERATOR}

As we mentioned, for bound states Eq. (1.13) can be replaced by
\begin{equation}
     \Phi_P (k^\prime) = -i \int {d^4 k \over (2 \pi)^2} \hat H_2(\eta_1 P
+ k^\prime) \hat H_2(\eta_2 P - k^\prime) \hat I(k^\prime , k;P) \Phi_P
(k)
\end{equation}
 ($P=p_1+p_2 $, $ k =\eta_2 p_1 -\eta_1 p_2$, $k^\prime =\eta_2
p_1^\prime -\eta_1 p_2^\prime $ and in the center of mass frame
$ {\bf P}=0, P_0 =\sqrt{s}, {\bf p}_1= -{\bf p}_2={\bf k}$,
 $ {\bf p}_1^\prime = -{\bf p}_2^\prime ={\bf k}^\prime $.\par
Let us recall the definitionof the instantaneous kernel
 given in Sec.1
 and consider the
approximation consisting in replacing in (5.1) $\hat{I} (k^\prime,
k;P)$ by $ \hat{I}_{\rm inst} ({\bf k}, {\bf k}^\prime)$
 and in  substituting
 $\hat{H}_{2 j}(p)$ with the free propagator ${-i \over p^2 +m^2}$.
 Further let us introduce
the reduced wave function
\begin{equation}
 \varphi_{P} ({\bf
k}^\prime) =
\sqrt{ 2 w_1({\bf k}) w_2({\bf k}^\prime ) \over w_1({\bf k}^\prime)
 + w_2({\bf k}^\prime )}
\int_{-\infty}^\infty d k_0^\prime \Phi_P ({ k}^\prime).
\label{eq:cinqdue}
\end{equation}
and integrate over $k_0$ and $k_0^\prime$ using
\begin{eqnarray}
\int dk_0^\prime & & {1\over (k_0^\prime + \eta_1 m_B)^2 -{\bf
k}^{2\prime} -m_1^2 + i \varepsilon}\,
{1\over (-k_0^\prime + \eta_2 m_B)^2 -{\bf
k}^{ 2\prime} -m_2^2 + i \varepsilon}= \nonumber \\
& & = - \pi i ( {w_1^\prime + w_2^\prime \over  w_1^\prime w_2^\prime
}) {1 \over m^2_B - (w_1^\prime + w_2^\prime )^2}
\end{eqnarray}
we have
 \begin{eqnarray}
&&
(w_1({\bf k}^\prime ) + w_2 ({\bf k}^\prime ))^2 \varphi_{m_B} ({\bf
k}^\prime ) +\nonumber \\
& & \quad \quad + \int {d^3 k \over (2 \pi )^3 }
 \sqrt{ w_1({\bf k}^\prime ) + w_2 ({\bf k}^\prime) \over 2
 w_1({\bf k}^\prime) w_2({\bf k}^\prime )} \hat{I}_{\rm inst}({\bf k}^\prime,
 {\bf k})  \sqrt{w_1({\bf k}) + w_2 ({\bf k})
\over 2 w_1({\bf k}) w_2({\bf k}) } \varphi_{m_B}({\bf k})=
 m^2_B \varphi_{m_B}({\bf k}^\prime)
\label{eq:cinqqua}
\end{eqnarray}
from which Eq.(1.21) immediately follows.\par
 The linear potential of Eq.(1.22) is then given by
\begin{equation}
   \langle {\bf k}^\prime \vert V \vert {\bf k} \rangle =
{1 \over w_1^\prime
   + w_2^\prime + w_1 + w_2} \langle {\bf k^\prime} \vert U \vert {\bf
k}\rangle + \dots
\label{eq:cinqsei}
\end{equation}
from which
 (neglecting kinematical factors becoming equal to one for $\vert {\bf
k}\vert = \vert {\bf k}^\prime \vert$)
we obtain Eq.(1.23).
By performing a ${1\over m^2}$ expansion on Eq. (1.23)
 we find  the  $q \bar{q} $ potential at ${1\over m^2}$ order
\begin{eqnarray}
\langle {\bf k}^\prime \vert V \vert {\bf k} \rangle & &  =
 -{4\over 3 } \alpha_s {1\over 2 \pi^2 {\bf Q}^2} - {\sigma\over \pi^2}
{1\over {\bf Q}^4}
  -{4\over 3} {\alpha_s \over 2 \pi^2} {1\over m_1 m_2 {\bf Q}^2}
 [{\bf q}^2 - {({\bf q} \cdot  {\bf Q})^2 \over {\bf Q}^2 } ]\nonumber \\
& & - {4\over 3} i \alpha_s \langle {\bf k}^\prime \vert
 {1\over 2 m_1} { {\bf \alpha_1} \cdot {\bf r} \over r^3 } -{1\over 2
m_2} { {\bf \alpha}_2 \cdot {\bf r}\over r^3 } \vert {\bf k} \rangle
+ {4\over 3} {\alpha_s \over 2 m_1 m_2} \varepsilon_{hkl} { k^\prime_l
+ k_l \over 2} ( \sigma_1^h +\sigma_2^h ) \langle {\bf k}^\prime \vert
{ r_k \over r^3 } \vert {\bf k}\rangle \nonumber \\
& & + {1\over 3} {\alpha_s \over m_1 m_2} \langle {\bf k}^\prime \vert
3 { r^h r^k \over r^5 } - {\delta^{hk}\over r^3} \vert {\bf k} \rangle
 \sigma_1^h \sigma_2^k + {4\over 3} {\alpha_s \over m_1 m_2}
 { 2 \pi \over 3} {\bf \sigma}_1 \cdot {\bf \sigma}_2 \nonumber \\
 & & - {\sigma \over 6} ({1\over m_1^2 } + {1\over m_2^2} - {1\over
m_1 m_2} ) \langle {\bf k}^\prime \vert {\bf q}_{\rm T}^2 r \vert {\bf
k}\rangle \nonumber \\
 & & - {\sigma \over 2} \varepsilon_{hkl} { k_l^\prime + k_l \over 2}
 ( {\sigma_1^h \over m_1^2 } + {\sigma_2^h \over m_2^2 })
 \langle {\bf k}^\prime \vert {r_k \over r} \vert {\bf k}\rangle
 + {\sigma i \over 2 } \langle {\bf k}^\prime \vert - {1\over m_1} {
{\bf \alpha_1 }\cdot {\bf r} \over r} + {1\over m_2 } {{ \bf \alpha_2 }
\cdot {\bf r} \over r} \vert {\bf k} \rangle +\dots
\end{eqnarray}
 with  $ {\bf q} ={ {\bf k} + {\bf k}^\prime \over 2}$,
 $ {\bf Q} = {\bf k}^\prime -{\bf k}$.\par
Passing to the coordinate representation we may also write
\begin{eqnarray}
V & & = {4\over 3} {\alpha_s \over r} + \sigma r
+ {4 \over 3} {\alpha_s \over m_1 m_2} \big \{ {1 \over 2 r}
 (\delta^{hk} + {r^h r^k \over r^2 }) q^h q^k \big \}_{\rm
W}\nonumber \\
& & - {4\over 3} i \alpha_s ( {1\over 2 m_1 } {\alpha_1 \cdot {\bf r}
 \over r^3 } - {1\over 2 m_2 } { \alpha_2 \cdot {\bf r} \over r^3})
+ {4\over 3} {\alpha_s \over 2 m_1 m_2} ({\bf \sigma}_1 + {\bf
\sigma}_2 ) \cdot ( {\bf r} \times {\bf q} )\nonumber \\
& & + {1\over 3} {\alpha_s \over m_1 m_2 } [ { 3 ( \sigma_1 \cdot {\bf
r} ) (\sigma_2 \cdot {\bf r}) \over r^5 } - {\sigma_1 \cdot \sigma_2
\over r^3  } ] + {4\over 3} {\alpha_s \over m_1 m_2 }
 { 2 \pi \over 3} (\sigma_1 \cdot \sigma_2) \delta^3({\bf r})\nonumber
\\
& & - {\sigma \over 6} ( {1\over m_1^2} + {1\over m_2^2} -{1\over m_1
m_2} ) \{ {\bf q}^2_{\rm T} r \}_{\rm W}\nonumber \\
& & - {\sigma \over 2} ( {\sigma_1 \over m_1^2 } +{\sigma_2 \over
m_2^2} ) \cdot ( {{\bf r} \over r} \times {\bf q} ) - {\sigma i
\over 2} [ {1\over m_1}  {\alpha_1 \cdot {\bf r} \over r} - {1\over
m_2} { \alpha_2 \cdot {\bf r} \over r} ]
\end{eqnarray}
where now ${\bf q}$ stands for the momentum operator.
Now, by  performing a Foldy--Wouthuysen  tranformation with  generator
\begin{equation}
S= {i \over 2 m_1} \alpha_1 \cdot {\bf q} - {i\over 2 m_2}
 \alpha_2 \cdot {\bf q}
\end{equation}
we end up with the ${1\over m^2}$ potential
 which coincides with the Wilson loop
potential \cite{BCP,BP94}:
\begin{eqnarray}
V && = -  \frac{4}{3}
 \frac{{\alpha}_s}{r} + \sigma r \nonumber \\
& &
  \frac{1}{2m_1m_2} \left\{
 \frac{4}{3} \frac{{\alpha}_s}{r}
(\delta^{hk} + \hat{r}^h \hat{r}^k) p_1^h p_2^k \right\}_{{\rm W}} -
\nonumber\\
{} & - & \sum_{j=1}^2 \frac{1}{6m_j^2} \{ \sigma \, r \,
 {\bf p}_{j{\rm T}}^2  \}_{{\rm W}} -
\frac{1}{6m_1m_2} \{ \sigma \, r \,
{\bf p}_{1{\rm T}} \cdot {\bf p}_{2{\rm T}} \}_{{\rm W}}
\nonumber \\
& &
\frac{1}{8} \left( \frac{1}{m_1^2}
 + \frac{1}{m_2^2} \right)
\nabla^2 \left( - \frac{4}{3} \frac{\alpha_s}{r} + \sigma r \right)
+
\nonumber\\
{} &+&  \frac{1}{2} \left(
 \frac{4}{3} \frac{\alpha_s}{r^3} -
\frac{\sigma}{r} \right) \left[ \frac{1}{m_1^2} {\bf S}_1 \cdot
( {\bf r} \times {\bf p}_1 ) - \frac{1}{m_2^2} {\bf S}_2 \cdot
( {\bf r} \times {\bf p}_2 ) \right] +
\nonumber\\
{} &+& \frac{1}{m_1m_2} \frac{4}{3} \frac{\alpha_s}{r^3} [ {\bf S}_2
\cdot ( {\bf r} \times {\bf p}_1 ) - {\bf S}_1 \cdot ( {\bf r}
\times {\bf p}_2 )] +
\nonumber\\
&+& \frac{1}{m_1m_2} \frac{4}{3} \alpha_s \left\{ \frac{1}{r^3}
\left[ \frac{3}{r^2} ({\bf S}_1 \cdot {\bf r})({\bf S}_2 \cdot
{\bf r}) - {\bf S}_1 \cdot {\bf S}_2 \right] +
\frac{8\pi}{3} \delta^3({\bf r}) {\bf S}_1 \cdot {\bf S}_2 \right\}
\> ,
\end{eqnarray}
with $\hat{{\bf r}} =({\bf r}/r)$ and the symbol $\{ \,\,\}_{\rm
W}$ stands for the Weyl ordering
prescription  among momentum  and position variables.

\section { CONCLUSIONS}
     In conclusion, under the assumption (1.2) and (1.3) for the evaluation
of the Wilson loop integral, we have derived a quark-antiquark
 Bethe-Salpeter (BS)
equation from QCD, extending a preceding result obtained for spinless quarks.
The assumptions are the same previously used for the derivation of a
semirelativistic heavy quark potential and  the technique is strictly
 similar.
The kernel is constructed as an expansion in $\alpha_{\rm s}$ and $\sigma
a^2$ and at the lowest order is given by equations (1.14)-(1.18).\par
The BS equation that has been obtained is a second order one,
analogous in some way to the iterated Dirac equation.
Correspondently,
by instantaneous approximation, an effective
square mass operator can be obtained from (1.19)
 which is given by (1.20) and
(1.21).\par
At the lowest order in $\alpha_s $  and $\sigma a^2$
 even a linear mass operator
can be  written with a potential $V$ given by (1.23). Neglecting
the spin dependent terms in $V$ the hamiltonian for the relativistic flux
tube model comes out. On the contrary by a ${1 \over m}$ expansion and
an appropriate Foldy-Wouthuysen transformation the ordinary semirelativistic
potential is reobtained.\par
In equation (1.13) or (1.19) a colour independent dressed  quark
propagator appears which is defined by equations (4.11) and (3.16). Notice
that only the perturbative expansion  gives contribution to this
quantity.\par
Few additional remarks are in order.\par
 First of all, notice that the result does not depend strictly from equation
(1.3) or (3.9) but from the possibility of writing the interaction term as
an integral on the  world lines of the quark and the antiquark,
 as evidenced in (3.10). Multiple
integrations of the same type would be admissible, as it occurres for
the perturbative contribution, but dependence of
the integrand on higher derivatives in the parameters $\tau_1$ and $\tau_2$
would not enable to carry on the argument. We have no actual justification
that $i \ln W$ is in general of the desired form, we observe however that this
quantity is obviously independent of the parametrization. For an example of
inclusion of higher order perturbative terms see Ref. [1].\par
A second point concerns the significance of the lowest order BS kernel we
have derived. As the analysis in terms of potentials show, the inclusion
of terms in $\alpha_{\rm s}$ is essential for an understanding of the fine
and the hyperfine structure. For what concerns the importance of $\sigma^2$
contributions an indication can be obtained considering the corresponding
terms in the relativistic tube flux model. Neglecting the coulombic terms
and in the equal mass case the c.m. hamiltonian for such model at the
$\sigma^2$ order can be written
\begin{eqnarray}
H_{\rm cm} & =& 2 \sqrt{m^2+{\bf q}^2} +{\sigma r\over 2}
 \Big [{\sqrt{m^2+{\bf q}^2}\over \vert {\bf q}_{\rm T}
\vert} {\rm arcsin}{\vert {\bf q}_{\rm T}\vert
 \over \sqrt{m^2+{\bf q}^2}}
+\sqrt{{m^2+ {\bf q}_{\rm r}^2\over m^2+{\bf q}^2}}\Big ]+\nonumber \\
 &+&{\sigma^2 r^2 \over 16  {\bf q}_{\rm T}^2} {m^2+{\bf q}_{\rm r}^2
\over \sqrt{m^2+{\bf q}^2}}
\Big  [ {\sqrt{m^2+{\bf q}^2} \over \vert {\bf q}_{\rm T}
\vert} {\rm arcsin} {\vert {\bf q}_{\rm T}\vert
 \over \sqrt{m^2+{\bf q}^2}}
-\sqrt{{m^2+ {\bf q}_{\rm r}^2\over m^2+ {\bf q}^2}}\Big ]^2
\end{eqnarray}
To better appreciate the relative magnitude of the two potential
  terms let us
consider e.g. the case of small ${\bf q}_{\rm T}$ (small angular momentum)
in which
the above equation becomes simply
   \begin{equation}
      H_{\rm cm} = 2 \sqrt {m^2 + {\bf q}^2 } + \sigma r + {\sigma^2 r^2
      \over 16 \sqrt {m^2 + {\bf q}^2}} \, .
  \end{equation}
Then, taking into account that $ a \sim 1/(\sigma m)^{1 \over 3}, \ q \sim
1/a $, and assuming typically
$ \sigma = 0.17 {\rm GeV}^2, \ m_{\rm u}=0.35 {\rm GeV} , \  m_{\rm c}=
1.7 {\rm GeV}, \  m_{\rm b}= 5{\rm GeV} $ we find that the last term in (6.2)
is of the order of the 5\%, 0.8\%, 0.2\% of the preceding one  for
the ${\rm u \bar u , \ c \bar c, \ b \bar b}$ systems respectively. This
would correspond to contributions to the mass of the meson of about 20,
2, 0.2 MeV.  The inclusion of the coulombic term would reduce $a$ and
improve the result. In the $ u\bar{u}$ case e.g. it would amount to
cut the above contribution by a factor 2.
Therefore only in this last case the $\sigma^2$ terms would be of
 any significance.
 \par
Finally  let us come to the problem
 of the type of confinement,
 which has been largely discussed in
the literature.  By this terminology it is usually meant  the
tentative  assumption of a BS (first order) confining kernel of the
instantaneous form
\begin{equation}
\hat{I}_{\rm conf}= (2 \pi )^3 \Gamma {\sigma \over \pi^2 } {1\over
 {\bf Q}^4},
\label{eq:istconc}
\end{equation}
 or even  the covariant counterpart of it
\begin{equation}
\hat{I}_{\rm conf} = - ( 2 \pi )^3 \Gamma {\sigma \over \pi^2 }
{1\over Q^4 },
\label{eq:covconc}
\end{equation}
where $\Gamma$ is a combination of Dirac matrices.
Typically   the cases $\Gamma =1$ (scalar
confinement), $ \Gamma = \gamma_1^0 \gamma_2^0 $ (vectorial
 confinement) or a combination of them
have been considered. \par
  Eq. (\ref{eq:covconc}) is
immediately ruled out by the fact that, even if formally it
corresponds to  (\ref{eq:istconc}) (by instantaneous approximation),
 actually, due to the strong infrared singularity, it gives results very
different from (\ref{eq:istconc}) \cite{retnos}.
As well known, Eq. (\ref{eq:istconc}) with $\Gamma=1$ was motivated by the
fact that it reproduces the static potential  $\sigma r$ and the spin
dependent potential as obtained in the Wilson loop  context. This
choice, however, gets both into phenomenological and theoretical
 difficulties:
\begin{itemize}
\begin{enumerate}
\item it gives a firts order velocity dependent relativistic
correction to the potential which differs from the Wilson loop
 one \cite{BCP,BP94} and does not seem to agree with the heavy
meson  data \cite{guptanoi},
\item  it does not reproduce straight line Regge trajectories
 \cite{durand,flux}.
\end{enumerate}
\end{itemize}
Complementary objections can be moved to (\ref{eq:istconc}) with
 $\Gamma = \gamma_1^0 \gamma_2^0 $.\par
On the contrary, even if we have not yet attempted calculations
directly with the kernel established in this paper, very encouraging
 results have been obtained in the context of the relativistic
 flux tube model \cite{flux}, of the dual QCD \cite{baker} and
 of the effective
 relativistic  hamiltonian \cite{sim}, formalisms that are all
 strictly related to our one. Therefore the complicated momentum
dependence appearing  in (1.16)-(1.17) seems essential to understand
both the light and the heavy meson phenomenology.

\appendix
\section{Appendix }

We want to prove Eq.(4.3).\par
 Let us first consider the confinement part
 and
 rewrite Eq. (3.3) as
\begin{equation}
 S_{\min} =
\int_{t_{\rm i}}^{t_{\rm f}} dt \int_0^1 ds \, {\cal S}(u)
\end{equation}
with
\begin{equation}
{\cal S}(u)=
\left[-
\left( \frac{\partial u^{\mu}}{\partial t} \frac{\partial u_{\mu}}
{\partial t} \right) \left( \frac{\partial u^{\mu}}{\partial s}
\frac{\partial u_{\mu}}{\partial s} \right) + \left(
\frac{\partial u^{\mu}}{\partial t} \frac{\partial u_{\mu}}
{\partial s} \right)^2 \right]^{\frac{1}{2}} \> .
\end{equation}
Being $x^\mu = u^\mu (s,t)$,
 the equation of the minimal surface $u^\mu$
enclosed by the loop  must be
 the solution of the Euler equations
\begin{equation}
\frac{\partial}{\partial s} \frac{\partial {\cal S}}{\partial
\left(\frac{\partial u^{\mu}}{\partial s} \right) } +
\frac{\partial}{\partial t} \frac{\partial {\cal S}}{\partial
\left(\frac{\partial u^{\mu}}{\partial t} \right) } =0
\end{equation}
satisfying the contour conditions $u^{\mu}(1,t)=z_1^{\mu}(\tau_1(t)),
\, u^{\mu}(0,t)=z_2^{\mu}(\tau_2(t))$. Then, considering an infinitesimal
variation of the world line of the quark 1, $z_1^{\mu}(t)
\longrightarrow z_1^{\mu}(t) + \delta z_1^{\mu}(t)$, even
$u^{\mu}(s,t)$ must change, $u^{\mu}(s,t)
\longrightarrow u^{\mu}(s,t) + \delta u^{\mu}(s,t)$ and one has
\begin{eqnarray}
\delta S_{\rm min}  &=&
\int_{t_{\rm i}}^{t_{\rm f}} dt \int_0^1 ds \, \left[  \frac{\partial
{\cal S}}{\partial \left(\frac{\partial u^{\mu}}{\partial s} \right) }
\frac{\partial}{\partial s} \delta u^{\mu} + \frac{\partial
{\cal S}}{\partial \left(\frac{\partial u^{\mu}}{\partial t} \right) }
\frac{\partial}{\partial t} \delta u^{\mu} \right] =
\nonumber\\
&=&  \int_{t_{\rm i}}^{t_{\rm f}} dt \, \left[ \frac{\partial
{\cal S}}{\partial \left(\frac{\partial u^{\mu}}{\partial s} \right) }
\delta u^{\mu} \right]_{s=1} \>
\end{eqnarray}
where
 $\delta z_1^{\mu}(t)$ is
assumed to vanish out of a small neighbourhood of a specific value of $t$.
Finally taking into account that
\begin{equation}
\delta u^{\mu}(1,t)= \delta z^{\mu}_1(t) \> , \qquad
\frac{\partial u^{\mu}(1,t)}{\partial t} =
\dot{z}^{\mu}_1(t)
\end{equation}
 one obtains
\begin{eqnarray}
\delta S_{\rm min} =
\int_{t_{\rm i}}^{t_{\rm f}} dt \, \frac{1}{[{\cal S}]_{s=1}}
\left[ - \dot{z}^2_1 \left( \frac{\partial u_{\nu}}{\partial s}
\right)_1 + \left( \frac{\partial u_{\mu}}{\partial s}
\right)_1 \dot{z}_1^{\mu} \dot{z}_{1\nu} \right] \delta z_1^{\nu} =
\nonumber\\
= \frac{1}{2}  \int_{t_{\rm i}}^{t_{\rm f}} dt \,
(dz^{\mu}_1 \delta z^{\nu}_1 - dz^{\nu}_1 \delta z^{\mu}_1)
\left[ \left( \frac{\partial u_{\mu}}
{\partial s} \right)_1 \dot{z}_{1\nu} - \left(
\frac{\partial u_{\nu}}{\partial s} \right)_1
\dot{z}_{1\mu} \right] \times
\nonumber\\
\times \left\{ - \dot{z}_1^2 \left( \frac{\partial u}
{\partial s} \right)_1^2 + \left[ \dot{z}_1 \cdot \left(
\frac{\partial u}{\partial s} \right)_1 \right]^2
\right\}^{-\frac{1}{2}}
\end{eqnarray}
and more explicitely
\begin{equation}
{\delta S_{\rm min} \over \delta S^{\mu \nu} (z_1)}=
 { ( {\partial u_\mu \over \partial s })_1 \dot{z}_{1\nu} -
 ({\partial u_\nu \over \partial s } )\dot{z}_{1 \mu}
\over [ - \dot{z}_1^2 ( {\partial u_\mu \over \partial s})_1
  + ( \dot{z}_1 ( {\partial u \over \partial s})_1 )^2 ]^{1\over 2}
}
\end{equation}
Then, in the straight line approximation we have
\begin{equation}
{\partial u_\mu \over \partial s } = z_{1\mu }-z_{2 \mu}= r_\mu
\end{equation}
and
\begin{equation}
{\partial S_{\rm min} \over \partial S^{\mu \nu} (z_1)} =
{ r_\mu \dot{z}_{1 \nu} - r_\nu \dot{z}_{1 \mu}\over
 [ - \dot{z}_1^2 r^2 + (\dot{z}_1 \cdot r )^2 ]^{1\over 2}}
\label{eq:resapp}
\end{equation}
Using Eq.(\ref{eq:resapp}) (having substituted the velocities
 with the momenta), Eqs.(3.11) and (3.6),
 we obtain the second term in (4.3).  \par
Let us come to the perturbative part. Consider a variation $z_1\to
 z_1 +\delta z_1 $, then
\begin{eqnarray}
\quad \quad
& & \delta_1 \int d\tau_1 \int d \tau_2  \dot{z}_1^\rho D_{\rho \sigma}
(z_1-z_2) \dot{z}_1^\sigma = \nonumber \\
& &= \int d \tau_1 \int d \tau_2 [ \delta \dot{z}_1^\rho D_{\rho
\sigma}(z_1-z_2) + \dot{z}_1^\rho \delta z_1^\nu \partial_\nu D_{\rho
\sigma} (z_1-z_2)] \dot{z}_2^\sigma =\nonumber \\
&&= \int \delta S^{\rho \nu} \int d\tau_2 [\partial_\nu D_{\rho \sigma}
 (z_1 -z_2) -\partial_\rho D_{\nu\sigma} (z_1- z_2 )]\dot{z}_2^\sigma
\end{eqnarray}
and so
\begin{equation}
{\delta \over \delta^{\mu \nu}(z_1)} \int d\tau_1\int d\tau_2
 p_1^\rho D_{\rho \sigma}(z_1-z_2) p_2^\sigma =
\int d\tau_2 (\delta_\mu^\rho \partial_{1\nu}-
\delta_\nu^\rho \partial_{1\mu} ) D_{\rho \sigma} (z_1-z_2) p_2^\sigma
\end{equation}
and then  we recover the first term in (4.3).

\end{document}